\documentclass[acmsmall,screen]{acmart}
\usepackage[utf8]{inputenc}
\usepackage[frozencache=true,cachedir=minted-cache]{minted} 
\usepackage{cleveref}

\setcitestyle{nosort} 

\newcommand{\hs}[1]{\mintinline{haskell}{#1}}

\setcopyright{rightsretained}
\acmPrice{}
\acmDOI{10.1145/3607849}
\acmYear{2023}
\copyrightyear{2023}
\acmSubmissionID{icfp23main-p55-p}
\acmJournal{PACMPL}
\acmVolume{7}
\acmNumber{ICFP}
\acmArticle{207}
\acmMonth{8}
\received{2023-03-01}
\received[accepted]{2023-06-27}

\ccsdesc[500]{Software and its engineering~Concurrent programming structures}
\ccsdesc[500]{Software and its engineering~Functional languages}

\keywords{Choreographic programming, freer monads}

\begin{document}

\title{HasChor: Functional Choreographic Programming for All (Functional Pearl)} 

\author{Gan Shen}
\orcid{0009-0006-0947-9531}
\affiliation{%
  \institution{University of California, Santa Cruz}
  \country{USA}}

\author{Shun Kashiwa}
\orcid{0009-0001-3665-0182}
\affiliation{%
  \institution{University of California, Santa Cruz}
  \country{USA}}

\author{Lindsey Kuper}
\orcid{0000-0002-1374-7715}
\affiliation{%
  \institution{University of California, Santa Cruz}
  \country{USA}}

\begin{abstract}
Choreographic programming is an emerging paradigm for programming distributed systems.
In choreographic programming, the programmer describes the behavior of the entire system as a single, unified program --- a \emph{choreography} --- which is then compiled to individual programs that run on each node, via a compilation step called endpoint projection.
We present a new model for functional choreographic programming where choreographies are expressed as computations in a monad.
Our model supports cutting-edge choreographic programming features that enable modularity and code reuse: in particular, it supports \emph{higher-order} choreographies, in which a choreography may be passed as an argument to another choreography, and \emph{location-polymorphic} choreographies, in which a choreography can abstract over nodes.
Our model is implemented in a Haskell library, \emph{HasChor}, which lets programmers write choreographic programs while using the rich Haskell ecosystem at no cost, bringing choreographic programming within reach of everyday Haskellers.
Moreover, thanks to Haskell's abstractions, the implementation of the HasChor library itself is concise and understandable, boiling down endpoint projection to its short and simple essence.
\end{abstract}

\maketitle

\section{Introduction}
\label{sec:intro}

A distributed system consists of a collection of \emph{nodes} that operate independently and communicate by message passing.
One of the challenges of programming distributed systems is the need to reason about the implicit \emph{global} behavior of the system while writing the explicit \emph{local} programs that actually run on each node.
While running independently, nodes must exchange messages in a carefully executed dance: every message sent from one node must be expected by its recipient, or we risk deadlock.

The emerging paradigm of \emph{choreographic programming}~\citep{montesi-dissertation,carbone-montesi-deadlock-freedom-by-design, cruz-filipe-montesi-core, cruz-filipe-chor-lambda, hirsch-garg-pirouette} helps to address this challenge by making the global behavior of the system \emph{explicit}.
In the choreographic paradigm, the programmer describes the behavior of a distributed system as a single, unified program: a \emph{choreography}.
One choreography is then compiled to multiple individual programs that run on each node, via a compilation step called \emph{endpoint projection}~\citep{carbone-cdl-epp-esop,carbone-cdl-epp}.
If endpoint projection is sound, the resulting distributed system enjoys a guarantee of \emph{deadlock freedom}~\citep{qiu-choreography-foundations,carbone-montesi-deadlock-freedom-by-design}: by construction, every message sent will be paired with a corresponding receive.
Furthermore, choreographies are amenable to whole-program analyses that can potentially rule out large classes of bugs.
However, choreographic programming --- and in particular \emph{functional} choreographic programming --- is still in its infancy, with the only existing functional choreographic language designs~\citep{hirsch-garg-pirouette, cruz-filipe-chor-lambda} so far lacking any practically usable implementation.

In this paper, inspired by the potential of functional choreographic programming, we present a programming model in which choreographies are expressed as computations in a monad.
We implement our model entirely as a Haskell library, which we call \emph{HasChor}.
Thanks to its embedding in Haskell, HasChor naturally supports cutting-edge choreographic programming features that enable a high level of abstraction.
HasChor programmers have access to all of Haskell's rich ecosystem, making functional choreographic programming viable for practical software development.
Furthermore, we find that Haskell's abstractions are a great fit for implementing the HasChor library itself, enabling a concise, understandable implementation of choreographic programming.

We make the following specific contributions:
\begin{itemize}

\item 
\emph{A monad for choreographic programming.}
Our main contribution is a new model for choreographic programming based on a monad and implemented as a Haskell library (\Cref{sec:api}).  
While choreographic programming (and functional choreographic programming) is not new, HasChor is to our knowledge the first practically usable implementation of functional choreographic programming (that is, implemented in a general-purpose programming language, rather than in a proof assistant or on paper), and the first to be based on a monad.
HasChor supports \emph{higher-order choreographies} and \emph{location-polymorphic choreographies}, both features that enable modularity and code reuse.

\item 
\emph{A case study for practical functional choreographic programming.}
As a case study, we use HasChor to implement a standard of the distributed systems literature: a replicated, in-memory key-value store (\Cref{sec:case-study}).
We build up the implementation in stages, showing how higher-order choreographies and location polymorphism enable a high level of abstraction in our key-value store implementation.
Our experience carrying out this case study suggests that HasChor is a practically usable implementation of functional choreographic programming.
It can be installed just like any Haskell library, compiled just like any Haskell program, and can use any Haskell tools for development and debugging, bringing choreographic programming within reach of everyday Haskellers.

\item 
\emph{An understandable implementation of choreographic programming.}
Finally, we contribute evidence that functional programming makes choreographic programming \emph{itself} straightforward to implement (\Cref{sec:impl}).
The core implementation of the HasChor library is less than 150 lines of code.\footnote{The optional HTTP backend adds another roughly 100 lines.}
We find that functional programming abstractions make it especially straightforward to implement endpoint projection, the central concept of choreographic programming.
In fact, HasChor's concise, simple implementation of endpoint projection helped \emph{us} grasp the essence of choreographic programming; we hope that it will do the same for readers.

\end{itemize}

We have published the HasChor implementation and a collection of example programs, including all of the examples from this paper, at \href{https://github.com/gshen42/HasChor}{github.com/gshen42/HasChor}.

\section{A tour of choreographic programming in HasChor}

In this section, we give a tour of choreographic programming with a series of examples and introduce its key ideas through the lens of HasChor.
We do not provide extensive explanations of language constructs in HasChor, but will point to sections in the rest of the paper where they are formally introduced.

\begin{figure}[t]
    \begin{minipage}[t]{0.45\textwidth}
        \begin{minted}[linenos]{haskell}
buyer :: Network IO (Maybe Day)
buyer = do
  send title "seller"
  price <- recv "seller"
  if price <= budget
  then do
    send True "seller"
    date <- recv "seller"
    return (Just date)
  else do
    send False "seller"
    return Nothing
        \end{minted}
    \end{minipage}
    \hfill
    \begin{minipage}[t]{0.45\textwidth}
        \begin{minted}[linenos]{haskell}
seller :: Network IO ()
seller = do
  title <- recv "buyer"
  send (priceOf title) "buyer"
  decision <- recv "buyer"
  if decision
  then do
    send (deliveryDate title) "buyer"
  else do
    return ()
        \end{minted}
    \end{minipage}
    \caption{The bookseller protocol implemented as individual programs}
    \label{fig:bookseller-individual}
\end{figure}

\begin{figure}[t]
    \centering
    \begin{minted}[linenos]{haskell}
bookseller :: Choreo IO (Maybe Day @ "buyer")
bookseller = do
  title'   <- (buyer, title) ~> seller
  price    <- seller `locally` \un -> return (priceOf (un title'))
  price'   <- (seller, price) ~> buyer  
  decision <- buyer `locally` \un -> return (un price' <= budget)
  
  cond (buyer, decision) \case
    True -> do
      date  <- seller `locally` \un -> return (deliveryDate (un title'))
      date' <- (seller, date) ~> buyer
      return (Just date')
    False -> do
      return Nothing
    \end{minted}
    \caption{The bookseller protocol implemented as a choreography}
    \label{fig:bookseller-choreo}
\end{figure}

\subsection{The Bug-Prone Bookseller}
\label{subsec:intro-bug-prone}

As an example of the kind of bug that choreographic programming is designed to prevent, consider a well-known example from the literature: the ``bookseller'' protocol~\citep{carbone-cdl-epp-esop,carbone-cdl-epp,honda-mpsts,w3c-cdl-primer}.
The first version of the protocol that we will consider involves an interaction between two participants: a seller and a (would-be) buyer.
The protocol begins with the buyer sending the title of a book they want to buy to the seller.
The seller replies with the book's price, and the buyer checks if the price is within their budget.
If the buyer can afford the book, they inform the seller and get back a delivery date; otherwise, they tell the seller they will not buy the book.

\Cref{fig:bookseller-individual} shows an implementation of the bookseller protocol as individual programs.
We call these programs \emph{network programs} and they are written in the \hs{Network} monad (\Cref{sec:network-monad}) provided by HasChor.
The \hs{buyer} has a particular \hs{title :: String} and a \hs{budget :: Int} in mind, and the \hs{seller} has \hs{priceOf :: String -> Int} and \hs{deliveryDate :: String -> Day} functions for looking up the price of a book and the date on which a book can be delivered.
Network programs interact with each other by sending messages with \hs{send} and receiving them with \hs{recv}.
The \hs{send} function takes as arguments a message of a serializable type (such as a book \hs{title} on line 3 of the \hs{buyer}, or a \hs{Bool} on lines 7 and 11 of the \hs{buyer}) and a destination location (such as \hs{"seller"} on lines 3, 7, and 11 of the \hs{buyer}), and transmits the message to the destination via some as-yet-unspecified transport mechanism.
The \hs{recv} function takes a location as its argument, blocks until a message has arrived from the specified location via the transport mechanism, and then returns the message.

Even in a simple protocol like this one, it is easy to introduce a bug that will lead to a deadlock.
For example, suppose that the implementor of \hs{buyer} forgets to write \hs{send True seller} on line 7.
Then both \hs{buyer} and \hs{seller} will be stuck: the \hs{seller} will wait forever on line 5 for a decision from the \hs{buyer}, while the \hs{buyer} will wait forever on line 8 for a date from the \hs{seller}.

\subsection{The Choreographic Approach}
\label{subsec:intro-choreographic-approach}

As a first example of choreographic programming in HasChor, \Cref{fig:bookseller-choreo} shows the implementation of the bookseller protocol as a single choreography.
The choreography expresses the interaction between locations (\Cref{subsec:api-locations}) --- in this case, buyer and seller --- from an objective, \emph{global} point of view, like the script of a play~\citep{giallorenzo-multiparty-languages}; in HasChor, choreographies are written in the \hs{Choreo} monad (\Cref{sec:choreo-monad,sec:choreo-impl}).
The \hs{bookseller} choreography returns a value of type \hs{Maybe Day @ "buyer"}, which is a located value (\Cref{subsec:api-loc-val,subsec:impl-located-values}) that represents a value of type \hs{Maybe Day} at the \hs{"buyer"} location.\footnote{Do not confuse the \hs{@} symbol with type application. Throughout this paper, the \hs{@} symbol always refers to located values.}

One of the hallmarks of choreographic programming is a single language construct that takes the place of both \hs{send} and \hs{recv}. 
For example, \hs{send title "seller"} (from line 3 of the \hs{buyer} in \Cref{fig:bookseller-individual}) and \hs{recv "buyer"} (from line 3 of the \hs{seller} in \Cref{fig:bookseller-individual}) are replaced in \Cref{fig:bookseller-choreo} by line 3 of the choreography:
\begin{minted}{haskell}
  title' <- (buyer, title) ~> seller
\end{minted}
where \hs{(~>)} is the HasChor construct that represents communication between two locations.
Here, \hs{title} represents the title sent by the buyer, while \hs{title'} represents the title received by the seller.
Locations can also perform local actions by using the \hs{locally} function, which takes as arguments a location and a local computation (in this case, an \hs{IO} computation) with access to a special \hs{un} function that can unwrap located values into normal values for use with other Haskell functions.
For example, on line 4 of \Cref{fig:bookseller-choreo}, the seller looks up the price of the book by doing a local computation. 

The choreography in \Cref{fig:bookseller-choreo} also illustrates the use of \hs{cond}, another HasChor construct.
The \hs{cond} language construct implements \emph{knowledge of choice}, a distinctive feature of choreographic languages~\citep{carbone-montesi-deadlock-freedom-by-design, hirsch-garg-pirouette, giallorenzo-choral}.
To understand the intuition for \hs{cond}, notice how in the programs in \Cref{fig:bookseller-individual}, the buyer must explicitly inform the seller of its decision to buy the book or not, by calling either \hs{send True seller} or \hs{send False seller}, with the corresponding \hs{recv} on line 5 of \hs{seller}.
This explicit synchronization is necessary in the non-choreographic implementation of the protocol to make the seller's \hs{send} (or lack thereof) of the book's delivery date match up with a \hs{recv} (or lack thereof) on the buyer's side.
In HasChor, on the other hand, the \hs{cond} language construct takes care of inserting the necessary synchronization \emph{automatically}.
We describe \hs{cond} in more detail, along with the \hs{Choreo} monad and the rest of the HasChor API, in \Cref{sec:api}.

\subsection{Higher-Order Choreographies and Location Polymorphism}
\label{subsec:intro-higher-order-polymorphic}

\begin{figure}[t]
    \begin{minted}[linenos, highlightlines={1, 3, 7, 17-19, 21-26}]{haskell}
bookseller :: (Int @ "buyer" -> Choreo IO (Bool @ "buyer")) 
           -> Choreo IO (Maybe Date @ "buyer")
bookseller mkDecision = do
  title'   <- (buyer, title) ~> seller
  price    <- seller `locally` \un -> return (priceOf (un t))
  price'   <- (seller, price) ~> buyer  
  decision <- mkDecision p
  
  cond (buyer, decision) \case
    True -> do
      date  <- seller `locally` \un -> return (deliveryDate (un t))
      date' <- (seller, date) ~> buyer
      return (Just date')
    False -> do
      return Nothing

mkDecision1 :: Int @ "buyer" -> Choreo IO (Bool @ "buyer")
mkDecision1 p = do
  buyer `locally` \un -> return (un p <= budget)

mkDecision2 :: Int @ "buyer" -> Choreo IO (Bool @ "buyer")
mkDecision2 p = do
  price''  <- (buyer, price') ~> buyer2
  contrib  <- buyer2 `locall` \un -> return (un price' / 2)
  contrib' <- (buyer2, contrib) ~> buyer
  buyer `locally` \un -> return (un p <= un contrib' + budget)
    \end{minted}
    \caption{A higher-order version of the bookseller choreography with changes to \Cref{fig:bookseller-choreo} highlighted}
    \label{fig:bookseller-higher-order}
\end{figure}

\begin{figure}[t]
    \begin{minted}[linenos, highlightlines={1-2}]{haskell}
bookseller :: Proxy a -> Choreo IO (Maybe Day @ a)
bookseller buyer = do
  title'   <- (buyer, title) ~> seller
  price    <- seller `locally` \un -> return (priceOf (un title'))
  price'   <- (seller, price) ~> buyer  
  decision <- buyer `locally` \un -> return (un price' <= budget)
  
  cond (buyer, decision) \case
    True -> do
      date  <- seller `locally` \un -> return (deliveryDate (un t))
      date' <- (seller, date) ~> buyer
      return (Just date)
    False -> do
      return Nothing
    \end{minted}
    \caption{A location-polymorphic version of the bookseller choreography with changes to \Cref{fig:bookseller-choreo} highlighted}
    \label{fig:bookseller-loc-poly}
\end{figure}

While the simple bookseller example suffices to illustrate the concept of a choreography, it is only the beginning of what is possible.
\citet{carbone-montesi-deadlock-freedom-by-design} use choreographies to implement a \emph{two-buyer} bookseller protocol, in which a second buyer, say, \hs{buyer2}, contributes half of the book's price to the budget.
We can implement this two-buyer protocol in HasChor by adding a little additional communication to the one-buyer choreography in \Cref{fig:bookseller-choreo}, for instance, by replacing line 6 of \Cref{fig:bookseller-choreo} with
\begin{minted}{haskell}
  price''  <- (buyer, price') ~> buyer2
  contrib  <- buyer2 `locally` \un -> return (un p' / 2)
  contrib' <- (buyer2, contrib) ~> buyer
  decision <- buyer `locally` \un -> return (un p <= un contrib' + budget)
\end{minted}
Indeed, it is a strength of choreographies that protocols involving three or more parties are straightforward to express.

However, we can do still better.
Recent work on choreographic programming proposes \emph{higher-order} choreographies~\citep{giallorenzo-choral,hirsch-garg-pirouette,cruz-filipe-chor-lambda}.
\citeauthor{hirsch-garg-pirouette} motivate the need for higher-order choreographies by pointing out that the one-buyer protocol and the two-buyer protocol share a common pattern that can be abstracted out: in each case, the buyer makes a decision to buy or not buy via some process, potentially involving communication.
With higher-order choreographies, this decision process can be implemented as a sub-choreography, which can then be passed as an argument to the main choreography, enabling code reuse.
\Cref{fig:bookseller-higher-order} shows the implementation of such a higher-order bookseller.
The \hs{bookseller} function takes a sub-choreography of type \hs{Int @ "buyer" -> Choreo IO (Bool @ "buyer")}.
Lines 17--19 of \Cref{fig:bookseller-higher-order} implement the single-buyer bookseller, equivalent to what we saw earlier in \Cref{fig:bookseller-choreo}, as the function \hs{mkDecision1}, while lines 21--26 of \Cref{fig:bookseller-higher-order} implement the two-buyer bookseller as the function \hs{mkDecision2}.

Another feature that raises the abstraction level of choreographic languages is \emph{location polymorphism}~\citep{giallorenzo-choral}.
As a motivating example, rather than implementing the bookseller protocol in a way that is specific to a \emph{particular} buyer, we might instead wish to implement a book-selling service to which an \emph{arbitrary} buyer may connect.
With location polymorphism, we can write a choreography that abstracts over the buyer as shown in \Cref{fig:bookseller-loc-poly}, again enabling code reuse and a higher level of abstraction.
Here, the \hs{bookseller} takes an argument of type \hs{Proxy a}\footnote{We use \hs{Proxy a} to represent a type-level location, which we will explain in more detail in \Cref{subsec:api-locations}.}, which could be any location.

Neither higher-order choreographies nor location polymorphism are new contributions of HasChor (see \Cref{sec:related} for a discussion of related work).
However, HasChor is to our knowledge the first practical \emph{functional} choreographic programming framework to support these features.
Still, we cannot claim too much credit: a happy consequence of our implementation approach is that both higher-order choreographies and location polymorphism ``just work'' in HasChor, because we inherit support for higher-order and polymorphic programming from our substrate of Haskell.

\subsection{Endpoint Projection}

The central concept of the choreographic programming paradigm --- and the technology that makes choreographies viable as a \emph{programming} language --- is \emph{endpoint projection} (EPP)~\citep{qiu-choreography-foundations,mendling-hafner-epp,carbone-cdl-epp-esop,carbone-cdl-epp}.
A choreography like that in \Cref{fig:bookseller-choreo} is already useful as a way of \emph{specifying} the global behavior of a protocol.
However, if we wish to actually \emph{run} the protocol in a distributed fashion, then we must have a way of extracting from the global choreography the individual network programs that will run at the buyer's and the seller's respective endpoints and make explicit calls to \hs{send} and \hs{recv}, like the programs in \Cref{fig:bookseller-individual}.
This is precisely what EPP does.
While EPP may sound complicated, Haskell's high-level abstractions --- in particular, the \emph{freer monad}~\citep{kiselyov-more-ext-effs} --- let us boil EPP down to its short and simple essence, with an implementation in just a few lines of code. 
We describe HasChor's implementation of EPP, along with other HasChor internals, in \Cref{sec:impl}.

Finally, to run the collection of projected programs produced by EPP, HasChor needs a message transport backend to actually implement sending and receiving, whether by HTTP, TCP, or messenger pigeon.
Here, again, the freer monad abstraction helps us: because freer monads separate the interface and implementation of an effectful computation, HasChor supports \emph{swappable backends} that implement a given interface.
Users of HasChor may implement their own backend or use the default HTTP backend that the HasChor framework provides.

\section{The HasChor API}
\label{sec:api}

In this section, we present the API of HasChor for writing and executing choreographies.
This section is oriented toward the user of HasChor and describes how to use the constructs provided by the library, while \Cref{sec:impl} provides an exposition of how these constructs are implemented.
HasChor can be viewed as an embedded domain-specific language for choreographic programming in Haskell.
The programming model it provides is typed, functional, higher-order, and polymorphic.
To fully support all these features, we require a number of language extensions from the Glasgow Haskell Compiler.  In particular, we use the \texttt{GHC2021}~\citep{GHC2021} set of language extensions, and additionally, we use \texttt{DataKinds} and \texttt{GADTs} extensively.

HasChor's API design is influenced by a variety of choreographic and \emph{multitier}~\citep{weisenburger-multitier-survey} languages, such as Pirouette~\citep{hirsch-garg-pirouette}, Choral~\citep{giallorenzo-choral}, ML5~\citep{murphy-ml5}, and ScalaLoci~\citep{weisenburger-scalaloci}; we discuss these and other related works in \Cref{sec:related}.

\subsection{Locations}
\label{subsec:api-locations}

Choreographic programming abstracts nodes in a distributed system as \emph{locations}.
These locations are treated atomically, and we assume location equality is decidable, so we can distinguish different locations.
In HasChor, we define a type alias \hs{LocTm} for locations, which we represent as \hs{String}s.  Because we also need locations to show up in types for \emph{located values} (which we describe next in \Cref{subsec:api-loc-val}), we also define type-level locations as \hs{Symbol}s (which are type-level \hs{String}s). 
\begin{minted}{haskell}
  type LocTm = String -- term-level location
  type LocTy = Symbol -- type-level location
\end{minted}
To provide a type-level location at the term level, we use the standard \hs{Proxy} datatype.
For example, the \hs{buyer} in the \hs{bookseller} choreography we saw in \Cref{fig:bookseller-choreo} is defined as a \hs{Proxy} of type \hs{Proxy "buyer"}:
\begin{minted}{haskell}
  buyer :: Proxy "buyer"
  buyer = Proxy -- a term-level proxy for a type-level location
\end{minted}

\subsection{Located Values}
\label{subsec:api-loc-val}

Since choreographic programming provides a global view of distributed programming, values at different locations will show up in the same choreography, which would make it possible for a location to access a value that doesn't reside on it, if we are not careful.
To avoid this problematic behavior, HasChor annotates each value with the location where it resides and ensures that only that location can access it.
We call these annotated values \emph{located values}, and write their types as \hs{a @ l}, which represents a value of type \hs{a} at location \hs{l}.

Located values are not immediately usable. To use a located value in a choreography, the value must first be ``unwrapped'' to a normal value by applying an \emph{unwrap function} of type \hs{forall a. a @ l -> a} to it.
Crucially, HasChor needs to ensure that only location \hs{l} is allowed to unwrap an \hs{a @ l} to \hs{a} and use it.
To accomplish this, we leave the definition of located values opaque to the user, and only provide access to the unwrap function in a safe manner through the \hs{Choreo} monad.

\subsection{The Choreo Monad} 
\label{sec:choreo-monad}

The programming model of HasChor is structured around a monad, \hs{Choreo}.
In HasChor, choreographies are computations of type \hs{Choreo m a}:
\begin{minted}{haskell}
  type Choreo m a 
  instance Functor (Choreo m)
  instance Applicative (Choreo m)
  instance Monad (Choreo m)
\end{minted}
\hs{Choreo} computations are parameterized by a user-supplied \emph{local monad} \hs{m}.
HasChor assumes that each location participating in a choreography of type \hs{Choreo m a} can run \emph{local computations} of type \hs{m a}.
The user can choose the local monad as they wish, with the only requirement being that the local monad needs to subsume \hs{IO}, i.e. being an instance of \hs{MonadIO}, as each node should be able to send and receive messages.
For example, the \hs{bookseller} choreography in \Cref{fig:bookseller-choreo} has type \hs{Choreo IO (Maybe Day)}.

As mentioned above, a \hs{Choreo} computation can be thought of as a program in an embedded DSL for choreographic programming in Haskell.
This embedded DSL supports three primitive language constructs: \hs{locally}, for carrying out local computations at a location; \hs{(~>)}, for communication between locations, and \hs{cond}, for choreographic conditionals.
We describe each of the language constructs in more detail below.

\subsubsection{Local computation}
\label{subsubsec:api-local-computation}

The \hs{locally} function is for doing local computation at a given location.
It takes a location \hs{l} and a local computation of type \hs{m a}, and returns a value of type \hs{a} located at \hs{l}:
\begin{minted}{haskell}
  locally :: Proxy l -> (Unwrap l -> m a) -> Choreo m (a @ l)
\end{minted}
The type of \hs{locally}'s second argument calls for some additional explanation.
The local computation is given an unwrap function of type \hs{Unwrap l}, which is an alias for \hs{forall a. a @ l -> a}.
The unwrap function allows the local computation to unwrap values located at \hs{l} in the context, but not values located at any other locations.
For example, the seller in the bookseller choreography in \Cref{fig:bookseller-choreo} uses the following code to look up the price of the book:
\begin{minted}{haskell}
  price <- seller `locally` \un -> return (priceOf (un t))
\end{minted}
The seller first unwraps the book title \hs{t} that it receives, and then calls \hs{priceOf} on it and returns the result.

\subsubsection{Communication}

A choreographic language must have a language construct for communication between a sender and a receiver.
The \hs{(~>)} function takes a pair of the sender's location and a value of type \hs{a} located at the sender, and a receiver's location, and returns a value of the same type \hs{a} but located at the receiver:
\begin{minted}{haskell}
  (~>) :: (Proxy l, a @ l) -> Proxy l' -> Choreo m (a @ l')
\end{minted}
For example, the seller in the bookseller choreography in \Cref{fig:bookseller-choreo} uses the following code to communicate the price of the book that it locally computed to the buyer:
\begin{minted}{haskell}
  price' <- (seller, price) ~> buyer
\end{minted}
Here, \hs{price'} has type \hs{Int @ "buyer"}, which represents the price the buyer receives.

For convenience, HasChor also provides a derived operation \hs{(~~>)} that combines \hs{locally} and \hs{(~>)}.  
The \hs{(~~>)} function carries out a local computation at a sender and then communicates the result to a receiver.
It takes as arguments a pair of the sender's location and a local computation of type \hs{m a}, and a receiver's location.
Like \hs{(~>)}, it returns a value of type \hs{a} located at the receiver.
\hs{(~~>)} has a straightforward implementation in terms of \hs{locally} and \hs{(~>)}:
\begin{minted}{haskell}
  (~~>) :: (Proxy l, Unwrap l -> m a) -> Proxy l' -> Choreo m (a @ l')
  (~~>) (l, c) l' = do
    x <- l `locally` c
    (l, x) ~> l' 
\end{minted}
For example, the seller in the bookseller protocol can use \hs{(~~>)} to combine looking up the price of the book and communicating that price to the buyer:
\begin{minted}{haskell}
  p <- (seller, \un -> return (priceOf (un t))) ~~> buyer
\end{minted}

\subsubsection{Conditionals}
\label{subsubsec:api-conditionals}

In choreographic programming, when one node in a system makes a choice such as taking one or another branch of a conditional, other nodes need to be informed of the choice in case it affects their communication pattern in the code that follows.
For example, in the bookseller protocol, the buyer's decision to buy or not buy the book must be communicated to the seller (as in the \hs{send True seller} and \hs{send False seller} calls in \hs{buyer} of \Cref{fig:bookseller-individual}).  
In a choreography, this would amount to writing a \hs{(~>)} expression in every branch of the conditional, which would be tedious for the programmer.
HasChor therefore provides a \hs{cond} language construct that inserts the necessary communication automatically.
\hs{cond} takes as its arguments a pair of a location and a \emph{scrutinee} at that location, and a function describing the follow-up choreographies depending on the scrutinee, and returns one of the follow-up choreographies.

\begin{minted}{haskell}
  cond :: (Proxy l, a @ l) -> (a -> Choreo m b) -> Choreo m b
\end{minted}
We have already seen an example use of \hs{cond} in the bookseller choreography, on lines 8--14 of \Cref{fig:bookseller-choreo}.
In that example, the scrutinee expression is \hs{decision}, which must be one of \hs{True} or \hs{False}.
While in the bookseller choreography the scrutinee happens to be of type \hs{Bool}, in general the scrutinee in a \hs{cond} expression may be of any type \hs{a}. 

HasChor's \hs{cond} makes a different design decision from most choreographic languages: typically~\citep{hirsch-garg-pirouette, giallorenzo-choral}, choreographic languages require the programmer to manually send synchronization messages in the branches of a conditional expression, to notify other locations about the decision that was made.
For example, in Pirouette~\citep{hirsch-garg-pirouette}, endpoint projection is undefined if a choreography neglects to include these synchronization messages.
In HasChor, on the other hand, the location where the conditional expression is evaluated will automatically broadcast the result to all locations.
This implementation strategy can make using conditionals a bit easier for users, at the cost of potentially adding unnecessary communication (for instance, by sending the result to locations whose behavior is not affected by it).
We discuss this trade-off in more detail in~\Cref{sec:related}.

\subsection{Running Choreographies}
\label{sec:api-running-choreographies}

To run a choreographic program in a distributed fashion, we must  project it to a collection of \emph{network programs} that run individually at each node.
The HasChor API provides a \hs{runChoreography} function that, given a choreography and a location name, projects the choreography to a network program at the specified location using endpoint projection, and then runs the network program:
\begin{minted}{haskell}
  runChoreography :: Backend config => config -> Choreo m a -> LocTm -> m a
\end{minted}
We defer further discussion of network programs and HasChor's implementation of endpoint projection to \Cref{sec:impl}.
As the type of \hs{runChoreography} shows, it also requires the user to provide a \emph{backend configuration} \hs{config}, which must be an instance of the \hs{Backend} type class.
To run a network program that is the result of endpoint projection, HasChor needs a \emph{message transport backend} to actually handle sending and receiving of messages.
A backend configuration specifies how locations are mapped to network hosts and ports, and provides a \hs{runNetwork} function that runs a network program at a specific location using some message transport mechanism.

While users are free to implement their own message transport backends, HasChor comes with an HTTP-based backend out of the box.
The HasChor API function \hs{mkHttpConfig} constructs an instance of \hs{Backend} that can be passed to \hs{runChoreography}.

Putting these pieces together, \Cref{fig:api-main-buyerseller} gives an example of using \hs{runChoreography} to project the bookseller choreography of \Cref{fig:bookseller-choreo} to a network program at each node, and then run the resulting network program at each node using HasChor's HTTP backend.

An important caveat about HasChor's support for higher-order choreographies is that it lacks \emph{separate compilation} for higher-order choreographies.  That is, in HasChor a higher-order choreography cannot be projected by itself, but must be applied to an argument choreography at projection time.  For example,  if we wished to project the higher-order choreography of \Cref{fig:bookseller-higher-order} to network programs, we would need to apply \hs{bookseller} to one of \hs{mkDecision1} or \hs{mkDecision2} before calling \hs{runChoreography}.  Other higher-order choreographic languages, such as Pirouette~\citep{hirsch-garg-pirouette} and Choral~\citep{giallorenzo-choral}, do not have this limitation.

\begin{figure}[t]
\begin{minted}{haskell}
main :: IO ()
main = do
  [loc] <- getArgs
  case loc of
    "buyer"   -> runChoreography cfg bookseller "buyer"
    "seller"  -> runChoreography cfg bookseller "seller"
    _         -> error ("Unknown location: " ++ loc)
  return ()
  where
    cfg = mkHttpConfig [ ("buyer",  ("alice.some-school.edu", 3000))
                       , ("seller", ("bookstore.haskell.org", 4000)) ]
\end{minted}
    \caption{Deploying the bookseller choreography of \Cref{fig:bookseller-choreo} with HasChor's HTTP backend}
    \label{fig:api-main-buyerseller}
\end{figure}

For testing and debugging purposes, HasChor users may also wish to run a \hs{Choreo} computation \emph{directly}, without the use of EPP.
To that end, HasChor provides a \hs{runChoreo} function, which has type \hs{Choreo m a -> m a}, and interprets a choreography as a non-distributed, single-threaded program.
We can view \hs{runChoreo} as giving a semantics to choreographies directly, rather than in terms of endpoint projection.
From a verification perspective, \hs{runChoreo} can be thought of as a \emph{specification} of how choreographies should behave, and correctness of endpoint projection becomes a question of whether the projected collection of network programs faithfully implements the specification.
We will discuss the implementation of \hs{runChoreo} further in our discussion of HasChor internals in \Cref{sec:impl}.

\section{Beyond Booksellers: A Replicated In-Memory Key-Value Store}
\label{sec:case-study}

In this section, we showcase the features of HasChor by using it to implement a standard of distributed systems: a replicated in-memory key-value store.
We build up the implementation in stages, starting first with a simple client-server architecture (\Cref{subsec:case-study-client-server}), and then a \emph{primary-backup} replication approach (\Cref{subsec:case-study-primary-backup}).
Next, we show how we can use HasChor's higher-order choreographies to \emph{abstract} over the previous two implementations (\Cref{subsec:case-study-higher-order-choreography}).
Finally, we show how we can use HasChor's location polymorphism to abstract out repeated code in the primary-backup implementation (\Cref{subsec:case-study-location-polymorphism}).

The full implementations of all our examples are available at \href{https://github.com/gshen42/HasChor}{github.com/gshen42/HasChor}.
Along with the key-value store examples that we describe in this section, these include an implementation of Diffie-Hellman key exchange~\citep{diffie-hellman-key-exchange}, a two-phase commit protocol~\citep{gray-2pc,lampson-sturgis-2pc}, and a distributed merge sort.

\subsection{A simple client-server key-value store}
\label{subsec:case-study-client-server}

\begin{figure}[t]
    \begin{minted}[linenos]{haskell}
type State = Map String String
data Request = Put String String | Get String
type Response = Maybe String
    
kvs :: Request @ "client" 
    -> IORef State @ "server" 
    -> Choreo IO (Response @ "client")
kvs request state = do
  request' <- (client, request) ~> server
  response <- server `locally` \un -> handleRequest (un request') (un state)
  (server, response) ~> client

handleRequest :: Request -> IORef State -> IO Response
handleRequest request state = case request of
  Put key value -> do
    modifyIORef state (Map.insert key value)
    return (Just value)
  Get key -> do
    state <- readIORef state
    return (Map.lookup key state)

mainChoreo :: Choreo IO ()
mainChoreo = do
  state <- server `locally` \_ -> newIORef Map.empty
  loop state
  where
    loop :: IORef State @ "server" -> Choreo IO ()
    loop state = do
      request <- client `locally` \_ -> readRequest
      response <- kvs request state
      client `locally` \un -> do putStrLn ("> " ++ show (un response))
      loop state
    \end{minted}
    \caption{The choreography for the client-server key-value store}
    \label{fig:code-client-server-kvs}
\end{figure}

\begin{figure}[t]
    \begin{minted}[linenos]{haskell}
main :: IO ()
main = do
  [loc] <- getArgs
  case loc of
    "client" -> runChoreography cfg mainChoreo "client"
    "server" -> runChoreography cfg mainChoreo "server"
    _        -> error ("Unknown location: " ++ loc)
  where 
    cfg = mkHttpConfig [ ("client", ("alice.some-school.edu", 3000))
                       , ("server", ("big-cloud-service.com", 4000)) ]
    \end{minted}
    \caption{Deploying the key-value store choreography with HasChor's HTTP backend}
    \label{fig:code-client-server-main}
\end{figure}

As a first step, we begin with a simple client-server architecture for our key-value store.
The client sends requests to the server, and the server handles requests and sends responses back to the client.
\Cref{fig:code-client-server-kvs} shows the HasChor implementation of the client-server key-value store.
The server stores pairs of \hs{String}s as a \hs{Map} inside a mutable \hs{IORef} as its \hs{State}.
It supports two kinds of \hs{Request}s: \hs{Put}, to set a given key-value pair, and \hs{Get}, to look up the value associated with a specified key.
The server uses \hs{Maybe String} as its response: for \hs{Put}, it sends back the value it put; for \hs{Get}, it sends back the corresponding value or \hs{Nothing} if it is not present.
The core of the implementation is the \hs{kvs} choreography, which describes one instance of the client-server interaction.
When \hs{kvs} is invoked, the client first sends the request to the server (line 9).
Then, the server calls \hs{handleRequest} to process the request, updates the server state as needed, and generates a response (line 10).
Finally, the server sends the response to the client (line 11).

The entry point for the key-value store is \hs{mainChoreo} (lines 22--32), which initializes the server state to the empty \hs{Map} and then enters an infinite loop. 
Inside the loop, the program reads a command from the client's terminal by calling \hs{readRequest} (omitted for brevity), then calls \hs{kvs} to process the request.
After the client prints the response, it goes back to the beginning of the loop and waits for the user to make another request.

To run the key-value store choreography, we use the \hs{main} function shown in \Cref{fig:code-client-server-main}.
The call to \hs{mkHttpConfig} sets up HasChor's HTTP backend, specifying a hostname and port number for each of the two locations (lines 9 and 10).
The program takes a location name as a command-line argument and starts the choreography.
Assuming the built executable is called \mintinline{shell}{kvs}, with \mintinline{shell}{kvs server} running, a client can interact with the key-value store from the command line:
\begin{minted}{shell}
  $ kvs client
  GET hello
  > Nothing
  PUT hello world
  > Just "world"
  GET hello
  > Just "world"
\end{minted}

\subsection{A replicated key-value store}
\label{subsec:case-study-primary-backup}

\begin{figure}[t]
    \begin{minted}[linenos, highlightlines={2, 6-11}]{haskell}
kvs :: Request @ "client" 
    -> (IORef State @ "primary", IORef State @ "backup")
    -> Choreo IO (Response @ "client")
kvs request (primarySt, backupSt) = do
  request' <- (client, request) ~> primary
  cond (primary, request') \case
    Put _ _ -> do
      req <- (primary, request') ~> backup
      ack <- (backup, \un -> handleRequest (un req) (un backupSt)) ~~> primary
      return ()
    Get _ -> return ()
  primary `locally` \un -> handleRequest (un primaryState) (un request')
  (primary, response) ~> client
    \end{minted}
    \caption{The choreography for the primary-backup key-value store with changes to \Cref{fig:code-client-server-kvs} highlighted}
    \label{fig:code-primary-backup}
\end{figure}

Now that we've seen a simple key-value store, let us consider how we can implement a ubiquitous feature of distributed systems: \emph{data replication}.
The implementation in \Cref{subsec:case-study-client-server} had only one server, so if the server fails, we lose all the stored data. 
Replication mitigates this risk by creating multiple copies of data and distributing them across multiple locations.

A classic replication approach is \emph{primary-backup} replication~\citep{alsberg-day-primary-backup}.
In primary-backup replication, we designate one node to be the \emph{primary}, with which clients interact, and all other nodes to be \emph{backup} nodes, with which the primary interacts.
To begin with, we will consider a simple primary-backup configuration with only one backup.
Both the primary and the backup store a full copy of the data.
The client sends requests to the primary, and in the case of a \hs{Get} request, the backup need not be involved at all; the primary can respond to the request using its own copy of the data.
In the case of a \hs{Put} request, the client forwards the request on to the backup, which applies the change to its own state and then sends back an acknowledgment to the primary.
After receiving the acknowledgment from the backup, the primary applies the update to its own state and finally sends a response back to the client. 
Therefore the client does not receive a response until \emph{both} replicas have applied the update.\footnote{We can contrast this approach with another common design in which the primary eagerly acknowledges the client's \hs{Put}s while asynchronously sending an update to the backups, which trades off strong consistency among replicas in exchange for lower request latency.}

\Cref{fig:code-primary-backup} shows an updated version of the \hs{kvs} choreography that uses primary-backup replication.\footnote{Other parts of the code are the same except that the configuration also needs to provide a mapping for \hs{backup}.}
As with the simple client-server setup that we saw in \Cref{fig:code-client-server-kvs}, \hs{kvs} takes a client request as its first argument.
However, because the choreography now needs to keep track of states on \hs{primary} and \hs{backup}, it takes a pair of replica states as its second argument.

Our use of \hs{cond} in \Cref{fig:code-primary-backup} illustrates HasChor's support for scrutinees of arbitrary type instead of just \hs{Bool}, as discussed in \Cref{subsubsec:api-conditionals}.
In the function passed to the \hs{cond} expression, we pattern-match on the variants of \hs{Request}.
If the request is a \hs{Put}, it must be forwarded to the backup, 
but \hs{Get} requests do not need to be passed to the backup, and are handled solely by the primary.

After the \hs{cond} has run its course, the primary handles the client request, be it \hs{Put} or \hs{Get}, by calling \hs{handleRequest}, and then communicates the response to the client.

\subsection{Abstracting over replication strategies: higher-order choreographies}
\label{subsec:case-study-higher-order-choreography}

\begin{figure}
    \begin{minted}[linenos]{haskell}
type ReplicationStrategy a = 
  Request @ "primary" -> a -> Choreo IO (Response @ "backup")

null :: ReplicationStrategy (IORef State @ "primary")
null request primarySt = 
  primary `locally` \un -> handleRequest (un request) (un primarySt)
    
primaryBackup :: 
  ReplicationStrategy (IORef State @ "primary", IORef State @ "backup")
primaryBackup request (primarySt, backupSt) =
  cond (primary, request) \case
    Put _ _ -> do
      req <- (primary, request') ~> backup
      ack <- (backup, \un -> handleRequest (un req) (un backupSt)) ~~> primary
      return ()
    Get _ -> return ()
    primary `locally` \un -> handleRequest (un request) (un primarySt)
    \end{minted}
    \caption{A type that characterizes replication strategy, with examples of no (\hs{null}) and primary-backup (\hs{primaryBackup}) replication strategies}
    \label{fig:null-primBack-replication-strategy}
\end{figure}

\begin{figure}
    \begin{minted}[linenos, highlightlines={2, 5}]{haskell}
kvs :: Request @ "client" 
    -> a -> ReplicationStrategy a
    -> Choreo IO (Response @ "client")
kvs request states replicationStrategy = do
  request' <- (client, request) ~> primary
  response <- replicationStrategy request' states
  (primary, response) ~> client        
    \end{minted}
    \caption{Key-value store choreography that takes a replication strategy, with changes to \Cref{fig:code-client-server-kvs} highlighted}
    \label{fig:code-kvs-replication-strategy}
\end{figure}

So far, we've seen two versions of our key-value store: one with no replication, and the other with primary-backup replication.
Comparing the \hs{kvs} choreographies in \Cref{fig:code-client-server-kvs,fig:code-primary-backup}, we see a common pattern: both choreographies take a request on the client and state(s) on the server, handle the request, and return the response on the client.
Indeed, from the \emph{client's} perspective, the behavior of the key-value store should be indistinguishable, regardless of what backups are being done or not done on the server’s side.

HasChor's support for higher-order choreographies lets us exploit this commonality and factor out the details of replication from \hs{kvs}.
To begin with, we define a \hs{ReplicationStrategy} type as shown in \Cref{fig:null-primBack-replication-strategy}, whose values describe the details of how the primary should replicate data;
since different replication strategies might keep track of different types of server state, we use a type variable \hs{a} to represent the type of the server state.

The simple client-server key-value store from \Cref{subsec:case-study-client-server} does \emph{no} replication, with the primary solely handling the request.
We define this strategy as \hs{null}, as shown in \Cref{fig:null-primBack-replication-strategy}.

On the other hand, the primary-backup key-value store from \Cref{subsec:case-study-primary-backup} needs the more sophisticated replication strategy \hs{primaryBackup} as shown in \Cref{fig:null-primBack-replication-strategy}, which forwards \hs{Put} requests to the backup.
Of course, further implementations of \hs{ReplicationStrategy} are possible, but for now, these two versions suffice for our example.

Finally, we modify \hs{kvs} to a function that takes a \hs{replicationStrategy} argument, and call \hs{replicationStrategy} to process each request, as shown in \Cref{fig:code-kvs-replication-strategy}.
This refactored version of the \hs{kvs} choreography describes the simple communication pattern between the server and the client, involving just two uses of \hs{(~>)}.
With the details of the replication strategy abstracted away, the new version of \hs{kvs} makes it easy to see what a key-value store \emph{does}, as far as clients are concerned: it accepts requests and produces responses.

When invoking the higher-order version \hs{kvs} from the entry point, we pass a concrete replication strategy.
That is, in the \hs{mainChoreo} function in \Cref{fig:code-client-server-kvs}, the call to \hs{kvs} on line 30 would become \hs{kvs request state null} if we don't want a backup, or \hs{kvs request state primaryBackup} if we do.
(In the latter case, we would also need to update \hs{mainChoreo} to initialize the server state to the empty \hs{Map} on both the primary and the backup.)

\subsection{Abstracting over backup nodes: location polymorphism}
\label{subsec:case-study-location-polymorphism}

\begin{figure}[t]
    \begin{minted}{haskell}
doBackup :: Proxy a -> Proxy b 
         -> Request @ a -> IORef State @ b 
         -> Choreo IO (Response @ a)
doBackup locA locB request state = do
  cond (locA, request) \case
    Put _ _ -> do
      request' <- (locA, request) ~> locB
      (locB, \un -> handleRequest (un request') (un state)) ~~> locA
      return ()
    Get _ -> return ()
    \end{minted}
    \caption{\hs{doBackup} location-polymorphic choreography}
    \label{fig:code-do-backup}
\end{figure}

\begin{figure}[t]
    \begin{minted}{haskell}
doubleBackup :: 
  ReplicationStrategy
    (IORef State @ "primary", IORef State @ "backup1", IORef State @ "backup2")
doubleBackup request (primarySt, backup1St, backup2St) = do
  doBackup primary backup1 request backup1St
  doBackup primary backup2 request backup2St
  primary `locally` \un -> handleRequest (un primarySt) (un request')
    \end{minted}
    \caption{A double-backup replication strategy}
    \label{fig:code-double-backup-replication-strategy}
\end{figure}

What's better than one backup?
How about two?
In production environments, one primary and one backup may not be enough to satisfy data durability requirements.
Indeed, distributed data storage systems such as Hadoop default to a replication factor of three~\citep{shvachko-hadoop}.
Therefore, we might wish to implement a \emph{double-backup} replication strategy for our key-value store.
This strategy is similar to the primary-backup approach of \Cref{subsec:case-study-higher-order-choreography}, but it replicates data to two backup locations to further improve durability.

A naive implementation of double-backup replication would repeat the part of the choreography that forwards the request from the primary to the backup, using the same code for forwarding to the second backup. 
However, with HasChor's location polymorphism, we can abstract away the shared logic into a location-agnostic choreography.

\Cref{fig:code-do-backup} defines a \hs{doBackup} choreography that involves two abstract locations: \hs{locA} and \hs{locB}. \hs{doBackup} takes a request at \hs{locA} and state at \hs{locB} as arguments, examines the type of the request at \hs{locA}, and handles the request at \hs{locB} if it is a \hs{Put} request.

The \hs{doubleBackup} function in \Cref{fig:code-double-backup-replication-strategy} implements our double-backup replication strategy using \hs{doBackup}.
It calls \hs{doBackup} twice with different backup locations, \hs{backup1} and \hs{backup2}.
We could also refactor \hs{primaryBackup} to call \hs{doBackup} once, and we can easily extend it to any number of backup locations. 
We could further generalize our approach to support replication strategies with different topologies, such as chain replication~\citep{van-renesse-chain-replication}.

\section{Implementation} \label{sec:impl}

In this section, we turn our attention to the implementation of the HasChor library itself.
The implementation is centered around two monads: \hs{Choreo}~(\Cref{sec:choreo-impl}), for choreographies, and \hs{Network}~(\Cref{sec:network-monad}), for network programs.
Both monads are implemented as a \emph{freer monad} instantiated with an effect signature that describes the effectful operations the monad provides, so we begin with a short primer on freer monads~(\Cref{sec:freer-monad}).
We also discuss our implementation of located values~(\Cref{subsec:impl-located-values}).
Finally, we present our implementation of endpoint projection~(\Cref{sec:epp}), the central concept of choreographic programming that links up \hs{Choreo} and \hs{Network}.

\subsection{Freer Monads}
\label{sec:freer-monad}

We start with the freer monad~\citep{kiselyov-more-ext-effs}, which is defined as follows:
\begin{minted}{haskell}
  data Freer f a where
    Return :: a -> Freer f a
    Do     :: f b -> (b -> Freer f a) -> Freer f a
\end{minted}
A freer monad \hs{Freer f a} represents an effectful computation that returns a result of type \hs{a}.
The parameter \hs{f :: * -> *} is an effect signature that defines the effectful operations allowed in the computation.
\hs{Return r} denotes a pure computation that returns a value \hs{r} of type \hs{a}.
\hs{Do eff k} denotes an effectful computation: the first argument \hs{eff :: f b} is the effect to perform, and it returns a result of type \hs{b};
the second argument \hs{k :: b -> Freer f a} is a continuation that specifies the rest of the computation given the result of the performed effect.

The first advantage of using freer monads is that they free us from defining boilerplate monad instances.
\hs{Freer f} is a monad given any effect signature \hs{f}:
\begin{minted}{haskell}
  instance Monad (Freer f) where
    return = Return

    (Return r) >>= f = f r
    (Do eff k) >>= f = Do eff (k >=> f)
\end{minted}
The monadic return simply corresponds to the \hs{Return} constructor.
The monadic bind operator directly applies the follow-up monadic computation to a pure computation or chains together the follow-up monadic computation with the continuation of the effectful computation.\footnote{The \hs{(>=>)} operator is the Kleisli composition and has type \hs{(a -> m b) -> (b -> m c) -> a -> m c}.}

The second advantage of using freer monads is that they separate the interface and implementation of effectful computations.
A freer monad by itself doesn't assign any meaning to effects; it merely accumulates them as a term. 
To interpret effects in a freer monad, we define an \hs{interpFreer} function that interprets the effects in a freer monad in terms of another monad:
\begin{minted}{haskell}
  interpFreer :: Monad g => (forall a. f a -> g a) -> Freer f a -> g a
  interpFreer handler (Return r) = return r
  interpFreer handler (Do eff k) = handler eff >>= interpFreer handler . k
\end{minted}
An effect handler, of type \hs{forall a. f a -> g a} for some monad \hs{g}, maps effects described by \hs{f} to monadic operations in \hs{g}.
\hs{interpFreer} takes such an effect handler and folds it over a freer monad:
for \hs{Return r}, we simply return \hs{r} in the monad without using the effect handler;
for \hs{Do eff k}, we use the effect handler \hs{handler} to interpret the effect \hs{eff}, then bind the result to the continuation \hs{k} and recursively interpret the result of running the continuation.

To use an effectful operation in the freer monad, we lift it into the monad by wrapping the effect in the \hs{Do} constructor with \hs{Return} as the continuation:
\begin{minted}{haskell}
  toFreer :: f a -> Freer f a
  toFreer eff = Do eff Return
\end{minted}
We use \hs{Freer} to define the \hs{Network} and \hs{Choreo} monads, as we'll see in the following sections.

\subsection{Located Values}
\label{subsec:impl-located-values}

Before discussing the implementation of the \hs{Choreo} monad and endpoint projection, let's first take a look at how located values are implemented, as we will use them to introduce those two concepts.
A located value \hs{a @ l} represents a value of type \hs{a} located at location \hs{l}:
\begin{minted}{haskell}
  data a @ (l :: LocTy) = Wrap a | Empty
\end{minted}
\hs{a @ l} has two constructors, \hs{Wrap} and \hs{Empty}.
\hs{Wrap} represents a located value from location \hs{l}'s point of view --- it's just a value of type \hs{a}.
\hs{Empty} represents a located value from locations other than \hs{l}'s point of view --- it's empty to them, and they should never try to access it.

Internally, we provide two functions \hs{wrap} and \hs{unwrap} to create and use a located value:
\begin{minted}{haskell}
  wrap :: a -> a @ l
  wrap = Wrap
  
  unwrap :: a @ l -> a
  unwrap (Wrap a) = a
  unwrap Empty    = error "Should never happen for a well-typed choreography!"
\end{minted}
The \hs{wrap} function simply calls the \hs{Wrap} constructor.
The \hs{unwrap} function unwraps a located value.
It is erroneous to unwrap an empty located value, as this represents a location accessing a value that doesn't reside on it.
As a result, the \hs{unwrap} function cannot be directly exposed to the user.
Instead, the user gets access to it through the argument of type \hs{Unwrap l} in the function passed to \hs{locally} (\Cref{subsubsec:api-local-computation}).
Additionally, when using \hs{unwrap} internally, we carefully arrange things to guarantee we never unwrap an empty value.

\subsection{The Choreo Monad} 
\label{sec:choreo-impl}

We are now ready to discuss the implementation of the \hs{Choreo} monad, which represents choreographies.
It is defined as a \hs{Freer} monad instantiated with the following \hs{ChoreoSig} effect signature:
\begin{minted}{haskell}
  type Unwrap l = forall a. a @ l -> a

  data ChoreoSig m a where
    Local :: Proxy l -> (Unwrap l -> m a) -> ChoreoSig m (a @ l)
    Comm  :: Proxy l -> a @ l -> Proxy l' -> ChoreoSig m (a @ l')
    Cond  :: Proxy l -> a @ l -> (a -> Choreo m b) -> ChoreoSig m b

  type Choreo m = Freer (ChoreoSig m)
\end{minted}
\hs{ChoreoSig} is parameterized by an underlying monad \hs{m} that represents local computations, i.e., computations that are run with \hs{locally} (\Cref{subsubsec:api-local-computation}).
Each constructor of \hs{ChoreoSig} corresponds to one of the three main language constructs we introduced in \Cref{sec:choreo-monad}.
In fact, the language constructs are merely these effects lifted into the \hs{Choreo} monad:
\begin{minted}{haskell}
  locally :: Proxy l -> (Unwrap l -> m a) -> Choreo m (a @ l)
  locally l m = toFreer (Local l m)
  
  (~>) :: (Proxy l, a @ l) -> Proxy l' -> Choreo m (a @ l')
  (~>) (l, a) l' = toFreer (Comm l a l')
  
  cond :: (Proxy l, a @ l) -> (a -> Choreo m b) -> Choreo m b
  cond (l, a) c = toFreer (Cond l a c)
\end{minted}

A \hs{Choreo} computation can be directly executed as a single-threaded local program with the \hs{runChoreo} function:
\begin{minted}{haskell}
  runChoreo :: Monad m => Choreo m a -> m a
  runChoreo = interpFreer handler
    where
      handler :: Monad m => ChoreoSig m a -> m a
      handler (Local _ m)  = wrap <$> m unwrap
      handler (Comm _ a _) = return $ (wrap (unwrap a))
      handler (Cond _ a c) = runChoreo $ c (unwrap a)
\end{minted}
The \hs{runChoreo} function interprets a \hs{Choreo} monad as a single monadic program, with the only intricacy being that we need to appropriately wrap and unwrap values.
\hs{Local l m} is interpreted as just running the local computation, \hs{m},
\hs{Comm s a r} is interpreted as directly returning the value being communicated, \hs{a};
and \hs{Cond l a c} is interpreted as directly applying the rest of the choreography \hs{c} to the scrutinee \hs{a}.

\subsection{The Network Monad} 
\label{sec:network-monad}

The \hs{Network} monad represents programs with explicit message sends and receives, which we call \emph{network programs}, and is the target of endpoint projection.
The \hs{Network} monad relies on a message transport backend to carry out message sends and receives, such as the HTTP backend discussed earlier in \Cref{sec:api-running-choreographies}.
Happily, because free monads separate the interface and implementation of an effectful computation, defining and using a new backend is as simple as defining and calling a new interpretation function.

The \hs{Network} monad is defined as a \hs{Freer} monad instantiated with the following \hs{NetworkSig} effect signature:\footnote{We assume each piece of data is an instance of \hs{Show} and \hs{Read} for (de)serialization and omit the instance declaration for clarity.}
\begin{minted}{haskell}
  data NetworkSig m a where
    Run   :: m a -> NetworkSig m a
    Send  :: a -> LocTm -> NetworkSig m ()
    Recv  :: LocTm -> NetworkSig m a
    BCast :: a -> NetworkSig m ()

  type Network m = Freer (NetworkSig m)
\end{minted}
Like \hs{ChoreoSig}, \hs{NetworkSig} is also parameterized by an underlying monad \hs{m} that represents local computations.
It has four constructors:
\hs{Run} corresponds to a local computation;
\hs{Send} and \hs{Recv} correspond to sending and receiving messages to and from a location, respectively;
and \hs{BCast} corresponds to broadcasting a message (that is, sending to all locations).  
As we will see shortly in \Cref{sec:epp}, we use broadcasting to implement endpoint projection for the \hs{cond} operation.
We lift the above effects into the \hs{Network} monad using \hs{toFreer}:
\begin{minted}{haskell}
  run :: m a -> Network m a
  run m = toFreer (Run m)
  
  send :: a -> LocTm -> Network m ()
  send a l = toFreer (Send a l)
  
  recv :: LocTm -> Network m a
  recv l = toFreer (Recv l)
  
  broadcast :: a -> Network m ()
  broadcast a = toFreer (BCast a)
\end{minted}

\hs{Network} can be implemented in a variety of ways with different message transport backends. HasChor supports this by providing a type class \hs{Backend}:
\begin{minted}{haskell}
  class Backend config where
    runNetwork :: config -> LocTm -> Network m a -> m a
\end{minted}
A message transport backend defines a configuration type that specifies how locations are mapped to network hosts and ports and a \hs{runNetwork} function that runs a network program at a specific location.
For example, in the provided HTTP backend, \hs{runNetwork} is implemented using the Servant~web~API library~\citep{servant}.
The configuration for the HTTP backend is a map from locations to hostnames and ports, as in \Cref{fig:api-main-buyerseller,fig:code-client-server-main}.
Each location runs a web server that provides an endpoint for clients to send it messages, and puts the messages into a buffer when it receives a message.
A \hs{Send a l} is interpreted as calling the endpoint at location \hs{l} with message \hs{a},
a \hs{Recv l} is interpreted as taking a message from location \hs{l} from the buffer,
and a \hs{Bcast a} is interpreted as a sequence of \hs{Send}s to all locations with message \hs{a}.

\subsection{Endpoint Projection}
\label{sec:epp}

We have finally arrived at endpoint projection (EPP), the interpretation of a choreography as the corresponding network program for a specific location.
In HasChor, EPP corresponds to interpreting a \hs{Choreo} program as a \hs{Network} program with respect to a term-level location.
The function \hs{epp} implements EPP in HasChor, and it is shown in \Cref{fig:epp}.\footnote{The function \hs{toLocTm :: forall (l :: LocTy). Proxy l -> LocTm} in \Cref{fig:epp} turns a type-level location to a term-level location and is defined as \hs{symbolVal} from the \texttt{GHC.TypeLits} module.}
The \hs{epp} function calls \hs{interpFreer} with a \hs{handler} that maps each effect in \hs{ChoreoSig} into a monadic action in \hs{Network}:
\begin{itemize}
    \item For effect \hs{Local l m}, if the location being projected to is the same as \hs{l}, then we \hs{run} the local computation \hs{m} given an \hs{unwrap} function and \hs{wrap} the result to a located value; otherwise we return a \hs{Empty} located value.
    \item For effect \hs{Comm s a r}, if the location being projected to is the same as the sender's location, \hs{s}, then we interpret it as a \hs{send} of \hs{a} to the receiver location \hs{r} and return an \hs{Empty} located value; if it's the same as the receiver's location \hs{r}, then we interpret it as a \hs{recv} from \hs{s} and \hs{wrap} the result to a located value; otherwise, we return a \hs{Empty} located value.
    \item For effect \hs{Cond l a c}, if the location being projected to is the same as \hs{l}, then this is the location who's making the decision, so it broadcasts the decision to all locations and then continues projecting the branches of the \hs{cond} expression; otherwise, the location \emph{receives} a decision and then continues projecting the branches with the received decision.
\end{itemize}
Since all effects return a located value, if the location being projected to owns the value, for example, on lines 6 and 10 of \hs{epp}, we use \hs{wrap <$>} to create a located value, indicating that the value resides on that location. Otherwise, on lines 7 and 11 of \hs{epp}, we return an \hs{Empty} value indicating that the value doesn't reside on that location.

The concise implementation of \hs{epp} closely resembles previous pen-and-paper presentations of endpoint projection~\citep[Fig. 9]{cruz-filipe-montesi-core}, while being executable code.
Consider the meaning of \hs{(~>)}, for instance: if you're the sender, it means \hs{send}; if you're the receiver, it means \hs{recv}; and if you're neither of those, it's a no-op.
This semantics is at the heart of choreographic programming, and our use of \hs{Freer} lets us express it cleanly in \hs{epp}, in a way that is completely decoupled from the actual message transport backend that implements \hs{send} and \hs{recv}.

Not only does this decoupling make the implementation of \hs{epp} elegant, it also makes HasChor developer-friendly by not mandating any particular choice of message transport mechanism, and instead enabling a well-defined way to plug in one's own transport layer, using the \hs{Backend} interface.
While our provided HTTP backend is intended for web programming, nothing else about HasChor is specific to the web setting.  
Alternative backends could make HasChor a viable option for programming in \emph{any} setting in which participants communicate by message passing.

\begin{figure}[t]
    \begin{minted}[linenos]{haskell}
  epp :: Choreo m a -> LocTm -> Network m a
  epp c l' = interpFreer handler c
    where
      handler :: ChoreoSig m a -> Network m a
      handler (Local l m)
        | toLocTm l == l' = wrap <$> run (m unwrap)
        | otherwise       = return Empty
      handler (Comm s a r)
        | toLocTm s == l' = send (unwrap a) (toLocTm r) >> return Empty
        | toLocTm r == l' = wrap <$> recv (toLocTm s)
        | otherwise       = return Empty
      handler (Cond l a c)
        | toLocTm l == l' = broadcast (unwrap a) >> epp (c (unwrap a)) l'
        | otherwise       = recv (toLocTm l) >>= \x -> epp (c x) l'
    \end{minted}
    \caption{Endpoint projection function \hs{epp}}
    \label{fig:epp}
\end{figure}

\section{Related work}
\label{sec:related}

Choreographies emerged in the mid-2000s in the context of web services.  The Web Services Choreography Model~\citep{w3c-choreography-model} and Web Services Choreography Description Language~\citep{w3c-cdl} specifications aimed to establish standards for global descriptions of the behavior of collections of communicating processes.  These standardization efforts took place with the participation of academic experts (particularly from the session types community), and they informed and inspired research on choreographies and endpoint projection~\citep{qiu-choreography-foundations,mendling-hafner-epp,carbone-cdl-epp-esop,carbone-cdl-epp,lanese-choreographies,mccarthy-krishnamurthi-crypto-epp,corin-projection-session-types}, laying the foundation for practical choreographic programming; \citet{montesi-book} gives an overview of these developments.

While choreographies as a \emph{specification} mechanism came earlier, \citet{carbone-montesi-deadlock-freedom-by-design} pioneered the concept of a choreographic \emph{programming} language with their Chor language. \citeauthor{carbone-montesi-deadlock-freedom-by-design} reason about the correctness of choreographies in terms of \emph{multiparty session types}~\citep{honda-mpsts}.  Endpoint projection is reminiscent of \citet{honda-mpsts}'s concept of projection of a global type to a local type, although at the term level rather than the type level.  The related literature on session types is too vast to summarize here, but \citet{huttel-session-types-survey} and \citet{ancona-behavioral-types-survey} provide good surveys.

\citet{hirsch-garg-pirouette}'s Pirouette language and \citet{cruz-filipe-chor-lambda}'s Chor$\lambda$ language are the first \emph{functional} choreographic programming languages. The focus of both these works is on the semantic foundations of functional choreographic programming, rather than on practically usable and deployable implementations.  Pirouette is implemented in the Coq proof assistant and stands out for having a fully mechanized proof of deadlock freedom.  Pirouette supports higher-order choreographic functions, but lacks support for location polymorphism.  Chor$\lambda$ supports higher-order choreographic functions and a limited form of location polymorphism, in which only top-level function definitions may be location-polymorphic.  In recent work, \citet{graversen-polychorlambda} present PolyChor$\lambda$, a successor to Chor$\lambda$ that supports full process polymorphism, i.e., process-polymorphic lambda expressions that are usable in arbitrary expression contexts.

Currently, the most fully-realized incarnation of choreographic programming in a practical language may be Choral~\citep{giallorenzo-choral}, an object-oriented language that extends Java with choreographic programming features.  In Choral, choreographies are objects, and so Choral supports higher-order choreographies in the sense that choreographic objects may contain fields that are themselves choreographic objects.  Choral also supports location polymorphism (which it calls \emph{role parameterization}); in fact, it was by porting examples of location-polymorphic Choral code to HasChor that we discovered that HasChor also enjoys support for location polymorphism.  Finally, Choral, like HasChor, is designed to be independent of any particular message transport mechanism.  HasChor's implementation on top of Haskell is in some ways analogous to Choral's implementation on top of Java, although, unlike Choral, HasChor is ``just a library'' and does not require a separate compiler.

As discussed in \Cref{subsubsec:api-conditionals}, the design of HasChor's \hs{cond} trades off efficiency for programmer convenience: rather than requiring programmers to manually write synchronization code in the branches of a conditional expression, HasChor inserts the needed communication automatically.
As \citet{dalla-preda-dynamic-choreographies} observe, ensuring that all participants in a choreography remain sufficiently ``aware of the evolution of the global computation'' seems to involve an efficiency/ease-of-use trade-off in choreographic languages in general.
One approach is to deem a choreography unprojectable if the programmer forgets to write the communication code necessary for knowledge of choice~\citep{carbone-cdl-epp-esop, carbone-cdl-epp, hirsch-garg-pirouette}.
On the other hand, HasChor's approach of inserting communication automatically is reminiscent of the choreographic language AIOCJ~\citep{dalla-preda-aiocj, dalla-preda-dynamic-choreographies}.

However, the HasChor approach is admittedly heavy-handed in that a \hs{cond} expression results in a broadcast to \emph{all} participants in the choreography, whether or not they actually need to know what choice was made.
Previous work on choreography \emph{amendment} and \emph{repair}~\citep{cruz-filipe-montesi-core, lanese-amending-choreographies, basu-choreography-repair} involves statically analyzing choreographies and inserting only the minimum amount of communication needed.
In particular, \citet{cruz-filipe-montesi-core}'s amendment analysis is based on \emph{merging}~\citep{carbone-cdl-epp}.
Because of HasChor's rather unconventional implementation approach that involves dynamic interpretation of freer monads, choreography amendment in the traditional sense would seem difficult to accomplish in HasChor, if not impossible.
It might still be possible, though, to improve the efficiency of HasChor's \hs{cond} by providing a way to annotate choreographies with the set of locations that participate in them.
We intend to investigate this idea further in future work.

Finally, choreographic programming is a close cousin of \emph{multitier} programming~\citep{cooper-links,serrano-hop,murphy-ml5,chlipala-urweb,serrano-prunet-hopjs,weisenburger-scalaloci} (see \citet{weisenburger-multitier-survey} for a comprehensive survey of the multitier programming literature).  Multitier programming emerged in response to the complexity of web programming in the early 2000s, which required programming the \emph{tiers} of an application in different languages (for instance, JavaScript for client-side code, Java for server-side code, and SQL for an underlying database tier).  In multitier programming, one uses a single, unified language to program all of the tiers in the same compilation unit, with a compiler taking care of splitting the program into deployable units in distinct languages --- a technique not unlike endpoint projection.  \citet{giallorenzo-multiparty-languages} offer a thoughtful exploration of the evident relationship between multitier and choreographic programming.  We posit that functional choreographic languages like HasChor could be a good platform for further investigation of this relationship; after all, many multitier languages, such as the pioneering multitier languages Links~\citep{cooper-links} and Hop~\citep{serrano-hop,serrano-prunet-hopjs}, \citet{murphy-ml5}'s ML5, \citet{chlipala-urweb}'s Ur/Web, and \citet{weisenburger-scalaloci}'s ScalaLoci, are distinctly functional in nature. ML5 and ScalaLoci in particular influenced the design of HasChor's API, especially our notion of located values.

\section{Conclusion and future work}

We presented HasChor, a framework for functional choreographic programming in Haskell.  Our model for choreographic programming is based on a monad, \hs{Choreo}, for expressing choreographies.  Our Haskell-based implementation means that we enjoy support for higher-order choreographies and location-polymorphic choreographies --- both recently-proposed features of choreographic languages --- essentially for free.  Furthermore, HasChor users have access to all of Haskell's libraries and tooling, bringing choreographic programming within reach of everyday Haskellers and bringing the power of Haskell within reach of choreographic programming.  Finally, HasChor is an \emph{understandable} and \emph{usable} implementation of choreographic programming: our implementation, based on the freer monad, makes endpoint projection straightforward and at the same time enables a programmer-friendly design of swappable network backends.  Although HasChor relies heavily on Haskell's monad abstraction and flexible type system, we believe the same approach could be applied to other languages as well.

HasChor is certainly no remedy for all the problems of distributed programming.  In particular, two of the biggest challenges of distributed programming are \emph{asynchrony}, in which messages take arbitrarily long to arrive at their destinations, and \emph{faults}, in which nodes may crash and messages may be lost entirely.  HasChor --- like other functional choreographic languages~\citep{hirsch-garg-pirouette,cruz-filipe-chor-lambda} --- does nothing in and of itself to address these difficulties.  In future work, we are interested in exploring the theory and practice of fault-tolerant and asynchronous choreographic programming, building on HasChor as a basis for experimentation.

Another avenue of future work is formalizing the semantics of HasChor, so that we can ultimately prove the correctness of endpoint projection and guarantee deadlock freedom.  Since \hs{Choreo} choreographies are directly runnable using \hs{runChoreo} as well as being projectable to network programs, it may be fruitful to view the direct semantics of \hs{Choreo} as a specification, and define the correctness of endpoint projection with respect to that specification.

\section*{Acknowledgements}

We thank Jonathan Castello, Andrew Hirsch, Fabrizio Montesi, Patrick Redmond, and the anonymous ICFP '23 reviewers, all of whom gave valuable feedback that improved this paper.
Additionally, Patrick Redmond contributed significantly to the implementation of HasChor's HTTP backend.

This material is based upon work supported by the National Science Foundation under Grant No. CCF-2145367. Any opinions, findings, and conclusions or recommendations expressed in this material are those of the author(s) and do not necessarily reflect the views of the National Science Foundation.

\bibliographystyle{ACM-Reference-Format}
\bibliography{references}

\end{document}